 \documentclass{article}

 \usepackage{english}

 \usepackage{amssymb}

 \usepackage{amstex}

 \newcommand{\beq}{\begin{equation}}
 \newcommand{\eeq}{\end{equation}}
 \newcommand{\al}{\alpha}
 \newcommand{\be}{\beta}
 \newcommand{\ga}{\gamma}
 \newcommand{\de}{\delta}

 \newcommand{\smallcap}{\mbox{\tiny $\cap$}}
\begin{document} 
\small

\vspace{1cm}
\noindent
  {\Huge\bf \hspace{-0.5cm} On the Replica Fourier Transfrom}
\vspace{1cm}

 \noindent
 {\large\bf Domenico M. CARLUCCI and Cirano DE DOMINICIS} 

 \vspace{0.5cm}
 {\bf

 \noindent
 D.M.C.\,\,\,: \, Department of Physics, Tokyo Institute of Technology ,\\ 
	 2-12-1 Oh-okayama, Meguro-ku, 
	 Tokyo 152 Japan  \\   
 E-mail:\,\,mimmo@@stat.phys.titech.ac.jp 
 \vspace{0.3cm}

 \noindent
 C.D.D.\,: \,S.Ph.T., CE Saclay, 91191 Gif-sur-Yvette, France; \\
 E-mail:\,\,cirano@@spht.saclay.cea.fr  \\[-0.2cm]

 }

\vspace{0.4cm}

\begin{center}
{\large\bf Abstrat}
\end{center}

The {\it Replica Fourier Transform} introduced previously is related to 
 the standard definition of Fourier transforms over a group. Its use is 
 illustrated by block-diagonalizing the eigenvalue equation of 
 a four-replica Parisi matrix.

\vspace{1cm}

\section{ The {\it replica} group} 
\vspace{0.3cm}

Given a Parisi {\it scenario} with $R$ replica symmetry breaking steps 
(Parisi,1980),   
each replica $\al=1,2,\dots,n$ can be univocally written  as a series  
of $R+1$ numbers 

        \begin{equation} 
	   \al \equiv (a_0,a_1,\dots, a_R) 
	   \hspace{1cm} 
	   \mbox{with} 
	   \hspace{1cm} 
	   a_j=0,1,\dots, p_j/p_{j+1} -1
	\end{equation}
where the $p_j$'s are the sizes of the Parisi sub-boxes 
(by definition $p_0\equiv n$ and $p_{R+1}\equiv 1$).
If we impose periodic conditions\footnote{Without loss of generality 
we identify the replica $n$ as the sequence $n\equiv (0,0,\dots,0)$.}, 
the set of these sequences has the structure of an additive group 
and precisely it can be represented as the following direct product 

	\begin{equation} 
	   \mathbb{F}_{p_0/p_1} 
	   \times
	   \mathbb{F}_{p_1/p_2} 
	   \times
	   \dots
	   \times
	   \mathbb{F}_{p_R/p_{R+1}}
	   \label{direct_product} 
	\end{equation}
where $\mathbb{F}_p$ is the additive group of the first $p$ integers 
{\it modulo} $p$. 	   

On this group, one can define a (co-)distance ({\it overlap} 
in spin glass jargon) 
by saying that two replicas, $\al\equiv(a_0,\dots,a_R)$ and 
$\be\equiv(b_0,\dots,b_R)$,  have overlap 
$\al \cap \be=t$ if $ (\al-\be) \equiv (0,0,\dots,0,a_t-b_t,\dots)$ 
with $a_t \neq b_t$ and the first $t$ entries equal to zero. 
It is easy to check that, given three replicas, 
their mutual distances satisfy the well-known ultrametric properties.

On the other hand, it will turn out to be useful to define 
another kind of distance ({\it co-overlap}): given two 
replicas, $\mu\equiv(m_0,\dots,m_R)$ 
and $\nu\equiv(n_0,\dots,n_R)$, we say that they have co-overlap 
$\mu \overset{*}{\cap} \nu = k$ 
if $(\mu -\nu)\equiv(\dots,m_{k-1}-n_{k-1},0,\dots,0)$ with 
$m_{k-1}\neq n_{k-1}$ 
and the last $R+1-k$ entries equal to zero. The co-overlap also 
satisfies ultrametric properties, with the difference that 
the base of the ultrametric isosceles triangle
is smaller than the other sides. 

Finally, since we deal with a direct product of groups, the {\it Haar} 
measure is simply the product of the {\it Haar} measures of each 
$\mathbb{F}_{p_j/p_{j+1}}$ 

	\begin{equation} 
	   \int d {\al} \,
	   \overset{\rm def}{=}
	   \, 
	   \sum_{a_0}\sum_{a_1} 
	   \,\dots\,
	   \sum_{a_R}
	\end{equation} 
and for later convenience we choose to keep it normalised to $p_0=n$. 

\vspace{0.5cm}

\section{ Character and Fourier Transform}
\vspace{0.3cm}

From (\ref{direct_product}), it follows immediately that 
the character $\chi(\al)$  of the replica group is simply the product 

	\begin{equation} 
	   \chi(\al) 
	   \,=\,
	   \prod_{j=0}^R 
	   \exp\left\{ 
		      2 \pi i \frac{a_j}{p_j/p_{j+1}}
	       \right\}
	       \label{character}
	\end{equation}	
for each $\al \equiv (a_0,a_1,\dots,a_r) \in {\cal A}$. The Fourier 
Transform of a generic function $f(\al)$ at {\it momentum} 
$\mu \in {\cal A}$ reads 

	\begin{equation} 
	   \widehat{f}(\mu)
	   \,=\,
	   \int d {\al }\, 
	   \chi 
	   \,
	   \left( \al \mu \right) \,
	   f(\al). 
	   \label{FT}
	\end{equation} 

In many interesting cases where one needs the Fourier Transform of a
function depending only on the overlap, {\it viz.} 
$ f(\al) \equiv f_{\al \smallcap n}$, the integration in (\ref{FT}) 
can be worked out shell by shell

	\begin{equation} 
	   \widehat{f}(\mu) 
	   =
	   \sum_{t=0}^{R+1}
	   \,
	   \left[
	         \int_{{\cal S}_t} 
	         d \al \,
	         \chi 
	         \left( \al \mu \right) 
	   \right]
	   \,
	   f_{\al\smallcap n=t}
	   \label{shellFT}
	\end{equation} 

where each integration is restricted to the subgroup 
$S_{t}\equiv \{ \al : \al \cap n=t \}$.  

In order to compute the Fourier Transform of a shell, let us consider 
the set ${\cal B}_t \equiv \{ \al : \al \cap n \geq t \}$, {\it i.e.} 
the complementary of a ball with radius $t$. 

Because of the relation 

	\begin{equation} 
	  \sum_{a_j=0}^{p_j/p_{j+1} -1} 
	  \exp\left\{
		     2 \pi i \frac{a_j}{p_j/p_{j+1}}
	       \right\}
	  \,=\,
	  \frac{p_j}{p_{j+1}} \,
	  \delta^{\mbox{\tiny Kr}}_{m_r 0}\,,
	\end{equation}
it is straightforward to verify that the Fourier Transform of  
$B_t$ at momentum $\mu$ ($\mu \overset{*}{\cap} n =k$) reads

	\begin{equation} 
	   \int_{{\cal B}_t}
	   d \al \,
	   \chi
               \left(
                     \al \mu 
	       \right)
	   =
	   \left\{ 
		  \begin{array}{ccc}
		  	  \overset{R}{\underset{j=t}{\prod}} 
		  \frac{p_j}{p_{j+1}}=p_t & & t \geq k \\[0.2cm] 
		   0  & & t < k 
		  \end{array} 
	  \right.
	\end{equation}

Consequently, the Fourier Transform of a shell is derived 
as the difference of two successive balls, thus giving

	\begin{equation} 
	  \int_{{\cal S}_t}
	  d \al \, 
	  \chi \left(\al \mu\right)
	  =
	  \int_{{\cal B}_t}
	  d \al \, 
	  \chi \left(\al \mu\right)
	  -
	  \int_{{\cal B}_{t+1}}
	  \kern-0.9em
	  d \al \, 
	  \chi \left(\al \mu\right)
	  = 
	   \left\{ 
		  \begin{array}{ccc} 
		   p_t-p_{t+1} & & t \geq k \\[0.2cm] 
		   - p_{t+1}   & & t = k-1 \\[0.2cm]
	           0           & & t < k-1
		  \end{array} 
	  \right.
 	\end{equation}

Note that, because of the translational invariance of the function $f$, 
its Fourier transform depends only on the co-overlap 
$\mu \overset{*}{\cap} n$.
By making use of the previous result and taking as null the 
values out of the range of definition, equation (\ref{shellFT}) 
becomes

	\begin{equation} 
	  \widehat{f}(\mu) 
	  =
	  \widehat{f}_{\mu \overset{*}{\smallcap} n=k}
	  =
	  \sum_{t=k}^{R+1} 
	  p_t\,
	  \left(
	 	f_t - f_{t-1} 
	  \right) \,.
	  \label{MPrft}
	\end{equation}
thus retrieving the {\it Replica Fourier Transform} introduced in 
(De Dominicis {\it et al.}, 1996. De Dominicis {\it et al.}, 1997). 
In the limit $R\to \infty$, (\ref{MPrft}) coincides with 
the transform defined by M\'ezard and Parisi (1991) for the 
continuum.

Besides this simple case, the same formalism can be extended 
to the situations where one needs to 
(Fourier) transform in the presence of a {\it passive} overlap, which 
is not to be summed over. For example, for a {\it 3}-index 
matrix $A^{\al \be; \gamma}$, parametrised in the usual way 

	\begin{equation} 
	  A^{\al\be;\gamma} 
	  \equiv A^r_t 
	  \hspace{1cm} 
	  \mbox{where} 
	  \hspace{1cm}
	  \begin{array}{ccl} 
	    r &=& \al \cap \be \\
	    t &=& \max( \al \cap \gamma; \be\cap\gamma)\,\,,
	  \end{array} 
	\end{equation}
the Fourier Transform with respect to the lower index at $\al\cap\be$ fixed, 
can be written 

	\begin{equation} 
	  \widehat{A}^r (\mu) 
	  =
	  \int d\zeta\,
	  \chi (\zeta \mu) 
	  A^{r}_{\zeta \smallcap n}
	  \hspace{1cm} 
	  \zeta
	  =\left\{
	          \begin{array}{ccc} 
	 	  \al- \ga & & \al \cap \ga \geq \be \cap \ga \\[0.1cm]
	          \be- \ga & & \mbox{otherwise} 
		  \end{array} 
	   \right. 
	   \label{3_index_FT}
	\end{equation}  
As above one can perform the integration shell by shell, with 
the slight difference that the occurrence of different 
ultrametric regions must be taken into account.
Indeed, one finds that the region $\zeta\cap n>r$ 
occurs twice and equation (\ref{3_index_FT}) becomes 

	\begin{equation} 
	   \widehat{A}^r(\mu) 
	   \,\equiv\,
	   \widehat{A}^r_{\mu \overset{*}{\smallcap} n=k}
	   \,=\,
	   \sum_{t=k}^{R+1} 
	   p_t^{(r)} 
	   \left(
		 A^{r}_t - A^{r}_{t-1}
	   \right)
	   \hspace{0.6cm} 
	   \mbox{where} 
	   \hspace{0.6cm} 
	   p_t^{(r)} 
	   =
	   \left\{
		  \begin{array}{ccl} 
		   p_t & & t \leq r \\
		  2 p_t & & t > r \,. 
		  \end{array}
	  \right.
	  \hspace{-1cm}
	 \end{equation}

In the same way, one can generalise this procedure for RFT in the presence 
of two passive overlaps {\it e.g.} for the longitudinal anomalous 
configuration of {\it 4}-index ultrametric matrices as shown 
in ref. (De Dominicis {\it et al.}, 1997). In other words, one can say that 
the Replica Fourier Transform has no ``perception'' 
of the ultrametric tree (the character does not depend on the 
matrix which it acts upon), but it extracts out the right 
weights according to the  passive overlaps involved. This property 
is useful when dealing with convolution products with 
several replicas, where it reduces drastically an otherwise 
quite tedious procedure. We illustrate it on the Longitudinal-Anomalous 
eigenvalue equation for a {\it 4-}replica Parisi matrix. It writes

	\begin{equation} 
	   \frac{1}{2}\,
	   \int d\rho \,d\sigma \, 
	   M^{\al \be; \rho \sigma} f^{\rho \sigma; \gamma} 
	   \,=\,
	   \lambda_A \,  
	   f^{\al \be; \gamma}\,
	   \label{eigenvalue} 
	\end{equation}
The left hamd side has contributions from the Replicon-like 
and Longitudinal Anomalous like configurations.  The former 
is diagonal, the latter is more delicate to treat. 
It can be written 

	\begin{equation}
	\frac{1}{2}\,
	\int d(\rho\kern-0.2em - \kern-0.2em \sigma)\, d\xi\, 
	  M^{\al\smallcap \be; \rho\smallcap \sigma}_{\zeta\smallcap \xi}
	  \,
	  f^{\rho\smallcap\sigma}_{\xi\smallcap n} 
           \,;
	   \hspace{0.7cm} 
	   \xi
	    =\left\{
		    \begin{array}{ccc}
		   \rho- \ga & & \rho \cap \ga \geq \sigma \cap \ga \\[0.1cm]
	           \sigma- \ga & & \mbox{otherwise} 
		  \end{array} 
	   \right. 
	   \hspace{-1cm}
	   \label{LA}
	\end{equation}
where $\zeta$ is defined in (\ref{3_index_FT}) and $\al\cap\be=r$ is fixed.
In (\ref{LA}) one recognizes a convolution which under Fourier 
Transform over $\zeta$  yields the product of $\widehat{f}$ and the 
Fourier Transformed of $M$ (called $K$). So we get

     \begin{equation}
        \sum_{s=0}^{R} 
	        \left\{
		       \delta_{r;s}^{\mbox{\rm\tiny Kr}}
		       K^{r;r}_{k;r+1} + 
		       \frac{1}{4} K^{r;s}_k \delta_s^{(k-1)}
		\right\} \widehat{f}^{s}_k=\lambda_A\, \widehat{f}^r_k \,.
      \end{equation}

where the $M$ matrix is now block-diagonalized 
(Temesv\'ari {\it et al.}, 1994. De Dominicis {\it et al.}, 1997.). 
The diagonal term 
is the contribution from the Replicon configuration, the weight 
$\de_s^{(k-1)}/2\equiv (p_s^{(k-1)}-p_{s+1}^{(k-1)})/2$ 
coming out from the sum over $(\rho -\sigma)$.

\vspace{1cm}

{\bf Acknowledgements.} One of us (D.M.C.) warmly thanks 
S.~Caracciolo and L.~Migliorini for stimulating discussions.
C.D.D. also thanks R.Balian for early discussions. 
D.M.C's research was supported by the JSPS under Grant 
No. P96215 and he would like to thank the statistical 
group for the kind ospitality during this work.

\clearpage

\begin{center} 
	{\bf References}
\end{center}

\vspace{0.2cm}\noindent
{\bf  \, De Dominicis C.and  Carlucci D. M., 1996.} 
Replica Fourier Transform and field theory of spin glass: a simple 
model, {\sl C.R. Acad. Sci. Paris}, 323, pp.263-269.

\vspace{0.2cm}\noindent
{\bf  \, De~Dominicis C., Carlucci D.M. and  Temesv\'ari~T., 1997.}
Replica Fourier Transform on ultrametric trees, and 
block-diagonalising multi-replica matrices, {\sl J. Phys. I France}, 7, 
pp. 105-115.

\vspace{0.5cm}\noindent
{\bf \, M\'ezard M. and Parisi G., 1991}
Replica field theory for random manifolds, 
{\sl J. Phys. I France}, 1, pp. 809-836.

\vspace{0.5cm}\noindent
{\bf \, Parisi G., 1980} 
A sequence of approximated solutions to the S-K model for spin glasses.
{\sl J. Phys. A Math. Gen.}, 13, pp. L115-L121.

\vspace{0.5cm}\noindent
{\bf  \, Parisi G., 1980} 
The order parameter for spin glasses: a function on the interval $0-1$.
{\sl J. Phys. A Math. Gen.}, 13, pp. 1101-1112.

\vspace{0.5cm}\noindent
{\bf  \, Parisi G., 1980} 
Magnetic properties of spin glasses in a new mean field theory.
{\sl J. Phys. A Math. Gen.}, 13, pp. 1887-1895.

\vspace{0.5cm}\noindent
{\bf  \, Temesv\'ari T., De~Dominicis C. and Kondor I., 1994}
Block-diagonalizing ultrametric matrices, 
{\sl J. Phys. A Math. Gen.}, 27, pp. 7569-7595.

\end{document}